\newcommand{\LANL}[3]{#3}

\newcommand{\FORMAT}[3]{\LANL{#1}{#2}{#3}}

\FORMAT{
  \documentclass[fleqn]{article}
  \usepackage{times,epsf,graphics,floatflt,wrapfig}
  \usepackage{nips99}
  }
{
  \documentclass[fleqn]{article}
  \usepackage{times,epsf,graphics,floatflt,wrapfig}
  }
{  
  \documentclass[fleqn]{article}
  \usepackage{times,epsf,graphics,floatflt,wrapfig}
  \usepackage[hyperindex,hyperfigures]{hyperref}
  \setlength{\oddsidemargin}{0.35in}
  \setlength{\textwidth}{5.75in}
  }

\FORMAT{\intextsep 0mm}{\intextsep 2mm}{\intextsep2mm}
\title{Entropy and Inference, Revisited}

\author{Ilya Nemenman,$^{1,2}$ Fariel Shafee,$^3$ and William Bialek$^{1,3}$
  \\
  $^1$NEC Research Institute, 4 Independence Way,
\FORMAT{}{\\}{}
  Princeton, New Jersey 08540\\
$^2$Institute for Theoretical Physics, University of California, Santa Barbara, CA 93106\\
  $^3$Department of Physics, Princeton University, 
\FORMAT{}{\\}{}
  Princeton, New
  Jersey 08544\\ {\it nemenman@itp.ucsb.edu,
    \{fshafee/wbialek\}@princeton.edu}}

\begin{document}

\maketitle

\begin{abstract}
We study properties of popular near--uniform (Dirichlet) priors for
learning undersampled probability distributions on discrete nonmetric
spaces and show that they lead to disastrous results.  However, an
Occam--style phase space argument expands the priors into their infinite
mixture and resolves most of the observed problems. This leads to a
surprisingly good estimator of entropies of discrete distributions.
\end{abstract}

\FORMAT{}{\newpage}{}

Learning a probability distribution from examples is one of the basic
problems in data analysis. Common practical approaches introduce a
family of parametric models, leading to questions about model
selection. In Bayesian inference, computing the total probability of
the data arising from a model involves an integration over parameter
space, and the resulting ``phase space volume'' automatically
discriminates against models with larger numbers of parameters---hence
the description of these volume terms as Occam factors
\cite{mackay,vijay}.  As we move from finite parameterizations to
models that are described by smooth functions, the integrals over
parameter space become functional integrals and methods from quantum
field theory allow us to do these integrals asymptotically; again the
volume in model space consistent with the data is larger for models
that are smoother and hence less complex \cite{bcs}.  Further, at
least under some conditions the relevant degree of smoothness can be
determined self--consistently from the data, so that we approach
something like a model independent method for learning a distribution
\cite{nb}.

The results emphasizing the importance of phase space factors in
learning prompt us to look back at a seemingly much simpler problem,
namely learning a distribution on a discrete, nonmetric space.  Here
the probability distribution is just a list of numbers $\{q_i\}$, $i =
1, 2, \cdots , K$, where $K$ is the number of bins or possibilities.
We do not assume any metric on the space, so that a priori there is no
reason to believe that any $q_i$ and $q_j$ should be similar.  The
task is to learn this distribution from a set of examples, which we
can describe as the number of times $n_i$ each possibility is observed
in a set of $N= \sum_{i=1}^K n_i$ samples. This problem arises in the
context of language, where the index $i$ might label words or phrases,
so that there is no natural way to place a metric on the space, nor is
it even clear that our intuitions about similarity are consistent with
the constraints of a metric space.  Similarly, in bioinformatics the
index $i$ might label n--mers of the the DNA or amino acid sequence,
and although most work in the field is based on metrics for sequence
comparison one might like an alternative approach that does not rest
on such assumptions.  In the analysis of neural responses, once we fix
our time resolution the response becomes a set of discrete ``words,''
and estimates of the information content in the response are
determined by the probability distribution on this discrete space.
What all of these examples have in common is that we often need to
draw some conclusions with data sets that are {\em not} in the
asymptotic limit $N \gg K$.  Thus, while we might use a large corpus
to sample the distribution of words in English by brute force
(reaching $N \gg K$ with $K$ the size of the vocabulary), we can
hardly do the same for three or four word phrases.

In models described by continuous functions, the infinite number of
``possibilities'' can never be overwhelmed by examples; one is saved
by the notion of smoothness. Is there some nonmetric analog of this
notion that we can apply in the discrete case?  Our intuition is that
information theoretic quantities may play this role.  If we have a
joint distribution of two variables, the analog of a smooth
distribution would be one which does not have too much mutual
information between these variables.  Even more simply, we might say that
smooth distributions have large entropy.  While the idea of ``maximum
entropy inference'' is common \cite{maxent}, the interplay between
constraints on the entropy and the volume in the space of models seems
not to have been considered.  As we shall explain, phase space factors
alone imply that seemingly sensible, more or less uniform priors on the
space of discrete probability distributions correspond to disastrously
singular prior hypotheses about the entropy of the underlying
distribution.  We argue that reliable inference outside the asymptotic
regime $N \gg K$ requires a more uniform prior on the entropy, and we
offer one way of doing this.  While many distributions are consistent
with the data when $N \leq K$, we provide empirical evidence that this
flattening of the entropic prior allows us to make surprisingly reliable
statements about the entropy itself in this regime.

At the risk of being pedantic, we state very explicitly what we mean by
uniform or nearly uniform priors on the space of distributions.
The natural ``uniform'' prior  is given by
\begin{equation}
  {\mathcal P}_{\rm u}(\{q_i\}) = {1\over Z_{\rm u}}\,\delta\left(
    1 - \sum_{i=1}^K q_i\right), \;\; Z_{\rm u} = \int_{\mathcal
    A}dq_1 dq_2 \cdots  dq_K 
  \,\delta\left( 1 - \sum_{i=1}^K q_i\right) 
\end{equation}
where the delta function imposes the normalization, $Z_{\rm u}$ is the
total volume in the space of models, and the integration domain
${\mathcal A}$ is such that each $q_i$ varies in the range $[0,1]$.
Note that, because of the normalization constraint, an {\em
  individual} $q_i$ chosen from this distribution in fact is not
uniformly distributed---this is also an example of phase space
effects, since in choosing one $q_i$ we constrain all the other
$\{q_{j\neq i}\}$. What we mean by uniformity is that all
distributions that obey the normalization constraint are equally
likely a priori.

Inference with this uniform prior is straightforward.  If our examples
come independently from $\{ q_i\}$, then we calculate the probability
of the model $\{ q_i\}$ with the usual Bayes rule: \footnote{If the data
are unordered,  extra combinatorial factors have to be included in $P(\{
n_i\} | \{ q_i\})$. However, these cancel  immediately in later
expressions.}
\begin{equation}
  P(\{ q_i\}| \{ n_i\} ) = \frac{P(\{ n_i\} | \{ q_i\})
    {\mathcal P}_{\rm u}(\{q_i\})}{P_{\rm u}(\{ n_i\})}, \;\;
  P(\{ n_i\} | \{ q_i\}) = \prod_{i=1}^K (q_i)^{n_i}.
\end{equation}
If we want the best estimate of the probability $q_i$ in the least
squares sense, then we should compute the conditional mean, and this
can be done exactly, so that \cite{ww,thesis}
\vspace{-0.5mm}
\begin{equation}
\langle q_i\rangle = {{n_i +1}\over{N+K}} .
\label{laprule}
\end{equation}
Thus we can think of inference with this uniform prior as setting
probabilities equal to the observed frequencies, but with an ``extra
count'' in every bin.  This sensible procedure was first introduced by
Laplace \cite{laplace}. It has the desirable property that events which have not been observed are not automatically assigned probability zero.

A natural generalization of these ideas is to consider priors that
have a power--law dependence on the probabilities, the so called Dirichlet family of priors:
\vspace{-0.5mm}
\begin{equation}
{\mathcal P}_\beta(\{q_i\}) = {1\over Z(\beta)}
\delta\left( 1 - \sum_{i=1}^K q_i\right)
\prod_{i=1}^K q_i^{\beta-1} \,,
\label{P(q)}
\end{equation}

It is interesting to see what typical distributions from these priors
look like. Even though different $q_i$'s are not independent random
variables due to the normalizing $\delta$--function, generation of
random distributions is still easy: one can show that if $q_i$'s are
generated successively (starting from $i=1$ and proceeding up to
$i=K$) from the Beta--distribution
\begin{equation}
  P(q_i) = B\left(\frac{q_i}{1-\sum_{j<i} q_j}; \beta, (K-i)\beta
  \right),\;\;\;\;  B\left(x; a,b \right) =
  \frac{x^{a-1}(1-x)^{b-1}}{B(a,b)}\,,
\label{betadistr}
\end{equation}

\begin{wrapfigure}{r}{63mm}
  \vspace{-0mm}
  \centerline{\epsfxsize=1.0\hsize\epsffile{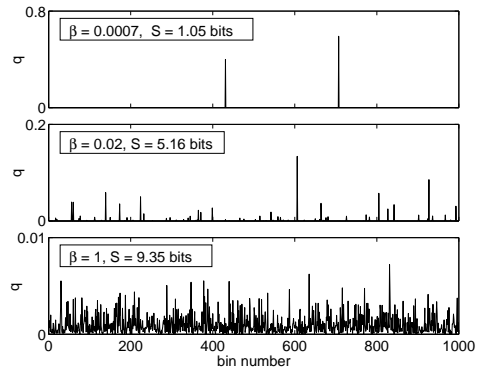}}  
  \vspace{-3.5mm}
  \caption{Typical distributions, $K=1000$.} 
  \FORMAT{\vspace{-1mm}}{}{}
  \label{example}
\end{wrapfigure}

\noindent then the probability of the whole sequence $\{q_i\}$ is ${\mathcal
  P}_{\beta}(\{q_i\})$.  Fig.~\ref{example} shows some typical
distributions generated this way. They represent different regions of
the range of possible entropies: low entropy ($\sim 1$ bit, where only
a few bins have observable probabilities), entropy in the middle of
the possible range, and entropy in the vicinity of the maximum,
$\log_2 K$.  When learning an unknown distribution, we usually have no
a priori reason to expect it to look like only one of these
possibilities, but choosing $\beta$ pretty much fixes allowed
``shapes.''  This will be a focal point of our discussion.

Even though distributions look different, inference with all priors
Eq.~(\ref{P(q)}) is similar \cite{ww,thesis}:
\begin{equation}
\langle q_i\rangle_\beta = {{n_i
+\beta}\over{N+\kappa}}\,,\;\;\;\; \kappa = K\beta.
\label{estim}
\end{equation}
This simple modification of the  Laplace's rule, Eq.~(\ref{laprule}),
which allows us to vary probability assigned to the outcomes not yet
seen, was first examined by Hardy and Lidstone \cite{hardy,lidstone}.
Together with the Laplace's formula, $\beta=1$, this family includes the
usual maximum likelihood estimator (MLE), $\beta \to 0$, that identifies
probabilities with frequencies, as well as the Jeffreys' or
Krichevsky--Trofimov (KT) estimator, $\beta=1/2$ \cite{jeffreys,kt,wst},
the Schurmann--Grassberger (SG) estimator, $\beta=1/K$ \cite{sg}, and
other popular choices.

To understand why inference in the family of priors defined by
Eq.~(\ref{P(q)}) is unreliable, consider the entropy of a distribution
drawn at random from this ensemble.  Ideally we would like to compute
this whole a priori distribution of entropies,
\begin{equation}
{\mathcal  P}_\beta (S) = \int dq_1  dq_2 \cdots dq_K \,
P_\beta(\{q_i\})
\,\delta\left[
S + \sum_{i =1}^K q_i\log_2 q_i \right] ,
\end{equation}
but this is quite difficult. However, as noted by Wolpert and Wolf
\cite{ww}, one can compute the moments of ${\mathcal P}_\beta (S)$
rather easily.  Transcribing their results to the present notation
(and correcting some small errors), we find:
\begin{eqnarray}
  \xi(\beta)  \equiv  \langle\, S [n_i =0]\, \rangle_\beta  &=& 
  \psi_0(\kappa+1) 
  -\psi_0(\beta+1) \, ,
  \label{Sap}
  \\
  \sigma^2(\beta) \equiv \langle \, (\delta S)^2  [n_i =0] \rangle_\beta
     &=& 
  \frac{\beta+1}{\kappa +
    1}\, \psi_1(\beta+1) -\psi_1(\kappa+1) \,,
  \label{dS2ap}
\end{eqnarray}
\vspace{-0.5mm}
where $\psi_m(x) = (d/dx)^{m+1} \log_2 \Gamma(x)$ are the polygamma
functions.

\begin{wrapfigure}{L}{63mm}
  \vspace{-1mm}
  \centerline{\epsfxsize=1.0\hsize\epsffile{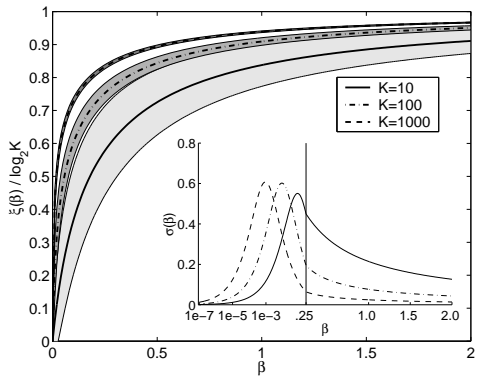}}
  \vspace{-4mm}
  \caption{$\xi(\beta) / \log_2
    K$ and $\sigma(\beta)$ as functions of $\beta$ and $K$; gray bands
    are the region of $\pm \sigma(\beta)$ around the mean. Note the
    transition from the logarithmic to the linear scale at
    $\beta=0.25$ in the insert.} 
  \FORMAT{\vspace{1mm}}{}{}
\label{Sapriori}
\end{wrapfigure}

This behavior of the moments is shown on Fig.~\ref{Sapriori}.  We are
faced with a striking observation: a priori distributions of entropies
in the power--law priors are extremely peaked for even moderately
large $K$. Indeed, as a simple analysis shows, their maximum standard
deviation of approximately 0.61 bits is attained at $\beta \approx
1/K$, where $\xi(\beta) \approx 1/\ln 2$ bits. This has to be compared
with the possible range of entropies, $[0, \log_2 K]$, which is
asymptotically large with $K$.  Even worse, for any fixed $\beta$ and
sufficiently large $K$, $\xi(\beta) = \log_2 K - O(K^0)$, and
$\sigma(\beta) \propto 1/\sqrt{\kappa}$. Similarly, if $K$ is large,
but $\kappa$ is small, then $\xi(\beta) \propto \kappa$, and
$\sigma(\beta) \propto \sqrt{\kappa}$.  This paints a lively picture:
varying $\beta$ between $0$ and $\infty$ results in a smooth variation
of $\xi$, the a priori expectation of the entropy, from $0$ to $S_{\rm
  max}= \log_2 K$.  Moreover, for large $K$, the standard deviation of
${\mathcal P}_{\beta} (S)$ is always negligible relative to the
possible range of entropies, and it is negligible even absolutely for
$\xi\gg 1$ ($\beta \gg 1/K$). Thus a seemingly innocent choice of the
prior, Eq.~(\ref{P(q)}), leads to a disaster: {\em fixing $\beta$
  specifies the entropy almost uniquely}.  Furthermore, the situation
persists even after we observe some data: {\em until the distribution
  is well sampled, our estimate of the entropy is dominated by the prior!}

Thus it is clear that all commonly used estimators mentioned above
have a problem. While they may or may not provide a reliable estimate
of the distribution $\{q_i\}$\footnote{In any case, the answer to
  this question depends mostly on the ``metric'' chosen to measure
  reliability. Minimization of bias, variance, or information cost
  (Kullback--Leibler divergence between the target distribution and
  the estimate) leads to very different ``best'' estimators.}, they
are definitely a poor tool to learn entropies.  Unfortunately, often
we are interested precisely in these entropies or similar
information--theoretic quantities, as in the examples (neural code,
language, and bio\-informatics) we briefly mentioned earlier.

Are the usual estimators really this bad? Consider this: for the MLE
($\beta=0$), Eqs.~(\ref{Sap}, \ref{dS2ap}) are formally wrong since it
is impossible to normalize ${\mathcal P}_0(\{q_i\})$.  However, the
prediction that ${\mathcal P}_0(S) = \delta(S)$ still holds. Indeed,
$S_{\rm ML}$, the entropy of the ML distribution, is zero even for
$N=1$, let alone for $N=0$. In general, it is well known that $S_{\rm
  ML}$ always underestimates the actual value of the entropy, and the
correction  \vspace{-0.5mm}
\begin{equation}
  S = S_{\rm ML} + \frac{K^*}{2N} + O \left( \frac{1}{N^2} \right) 
  \label{corr}
\end{equation}
\vspace{-0.5mm} is usually used (cf.~\cite{sg}).  Here we must set
$K^*=K-1$ to have an asymptotically correct result.  Unfortunately in
an undersampled regime, $N \ll K$, this is a disaster. To alleviate
the problem, different authors suggested to determine the dependence
$K^*=K^*(K)$ by various (rather ad hoc) empirical \cite{srrb} or
pseudo--Bayesian techniques \cite{pt}.  However, then there is no
principled way to estimate both the residual bias and the error of the
estimator.

\begin{wrapfigure}{r}{63mm}
  \vspace{-1mm}
  \centerline{\epsfxsize=1.0\hsize\epsffile{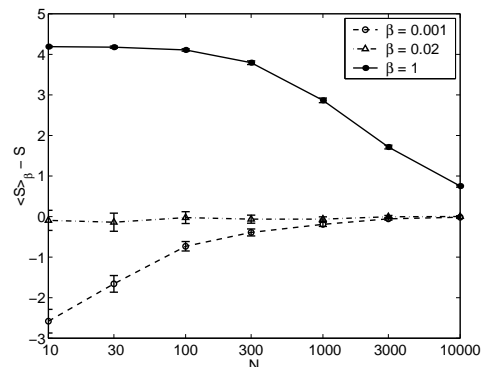}}
  \vspace{-5mm}
  \caption{Learning the $\beta=0.02$ distribution from  Fig.~\ref{example}
    with $\beta=0.001, 0.02, 1$. The actual error of the estimators is
    plotted; the error bars are the standard deviations of the
    posteriors. The ``wrong'' estimators are very certain but
    nonetheless incorrect.}  
  \FORMAT{\vspace{-2mm}}{}{\vspace{-2mm} }
  \label{fixedbeta}
\end{wrapfigure}

The situation is even worse for the Laplace's rule, $\beta=1$. We were
unable to find any results in the literature that would show a clear
understanding of the effects of the prior on the entropy estimate,
$S_{\rm L}$.  And these effects are enormous: the a priori
distribution of the entropy has $\sigma(1) \sim 1/\sqrt{K}$ and is
almost $\delta$-like. This translates into a very certain, but
nonetheless possibly wrong, estimate of the entropy. We believe that
this type of error (cf.~Fig.~\ref{fixedbeta}) has been overlooked in
some previous literature.

The Schurmann--Grassberger estimator, $\beta=1/K$, deserves a special
attention. The variance of ${\mathcal P}_{\beta}(S)$ is maximized near
this value of $\beta$ (cf.~Fig.~\ref{Sapriori}).  Thus the SG
estimator results in the most uniform a priori expectation of $S$
possible for the power--law priors, and consequently in the least
bias. We suspect that this feature is responsible for a remark in
Ref.~\cite{sg} that this $\beta$ was empirically the best for studying
printed texts. But even the SG estimator is flawed: it is biased
towards (roughly) $1/\ln 2$, and it is still a priori rather narrow.

Summarizing, we conclude that simple power--law priors,
Eq.~(\ref{P(q)}), must not be used to learn entropies when there is no
strong a priori knowledge to back them up. On the other hand, they are
the only priors we know of that allow to calculate $\langle q_i
\rangle$, $\langle S \rangle$, $\langle \chi^2 \rangle$, \dots exactly
\cite{ww}. Is there a way to resolve the problem of peakedness of
${\mathcal P}_{\beta}(S)$ without throwing away their analytical ease?
One approach would be to use $ {\mathcal P}^{\rm
  flat}_{\beta}(\{q_i\}) = \frac{{\mathcal P}_{\beta}(\{q_i\})
  }{{\mathcal P}_{\beta}(S[q_i])} \; {\mathcal P}^{\rm
  actual}(S[q_i])\,$ as a prior on $\{q_i\}$. This has a feature that
the a priori distribution of $S$ deviates from uniformity only due to
our actual knowledge ${\mathcal P}^{\rm actual} (S[q_i])$, but not in
the way ${\mathcal P}_{\beta}(S)$ does.  However, as we already
mentioned, ${\mathcal P}_{\beta}(S[q_i])$ is yet to be calculated.

Another way to a flat prior is to write ${\mathcal P}(S) = 1 = \int
\delta(S - \xi) d \xi$. If we find a family of priors ${\mathcal
  P}(\{q_i\}, {\rm parameters})$ that result in a $\delta$-function
over $S$, and if changing the parameters moves the peak across the
whole range of entropies uniformly, we may be able to use this.
Luckily, ${\mathcal P}_{\beta}(S)$ is almost a
$\delta$-function!~\footnote{The approximation becomes not so good as
  $\beta \to 0$ since $\sigma(\beta)$ becomes $O(1)$ before dropping
  to zero.  Even worse, ${\mathcal P}_{\beta}(S)$ is skewed at small
  $\beta$. This accumulates an extra weight at $S=0$.  Our approach to
  dealing with these problems is to ignore them while the posterior
  integrals are dominated by $\beta$'s that are far away from zero.
  This was always the case in our simulations, but is an open
question for the analysis of real data.} In addition, changing
$\beta$ results in changing $\xi(\beta) = \langle\, S [n_i=0] \,
\rangle_\beta$ across the whole range $[0, \log_2 K$]. So we may hope
that the prior \footnote{Priors that are formed as weighted sums of the
different members of the Dirichlet family are usually called {\em
Dirichlet mixture priors}. They have been used to estimate probability
distributions of, for example, protein sequences \cite{mixt}.
Equation (\ref{Pflat}), an {\em infinite} mixture, is a further
generalization, and, to our knowledge, it has not been studied before.}
\begin{equation}
{\mathcal P} (\{q_i\};\beta) = {1\over Z}\,
\delta\left( 1 - \sum_{i=1}^K q_i\right)
\prod_{i=1}^K q_i^{\beta-1} \frac{d \xi(\beta)}{d\beta} \,{\mathcal P}(\beta)
\label{Pflat}
\end{equation}
may do the trick and estimate entropy reliably even for small $N$, and
even for distributions that are atypical for any one $\beta$. We have less
reason, however, to expect that this will give an equally reliable
estimator of the atypical distributions themselves.$^2$ Note the term $d\xi/d\beta$  in Eq.~(\ref{Pflat}). It is there because $\xi$, not $\beta$, measures the position of the entropy density peak.

Inference with the prior, Eq.~(\ref{Pflat}), involves additional
averaging over $\beta$ (or, equivalently, $\xi$), but is nevertheless
straightforward. The a posteriori moments of the entropy are
\begin{eqnarray}
  \widehat{S^m} &=& \frac{\int d\xi\, 
    \rho(\xi,\{n_i\}) \langle\, S^m [n_i]\, \rangle_{\beta(\xi)}}
  {\int d\xi\, \rho(\xi,[n_i])}\,,\;\;\;\mbox{where}
  \label{Shat}
  \\
  \rho(\xi, [n_i]) &=& {\mathcal P}\left(\beta\left(\xi\right)\right)
  \frac{\Gamma(\kappa(\xi))}{\Gamma(N+\kappa(\xi))}\,
  \prod_{i=1}^K \frac{\Gamma(n_i+\beta(\xi))}{\Gamma(\beta(\xi))}\,.
  \label{rho}
\end{eqnarray}
Here the moments $\langle\, S^m [n_i]\, \rangle_{\beta(\xi)}$ are
calculated at fixed $\beta$ according to the (corrected) formulas of
Wolpert and Wolf \cite{ww}.  We can view this inference scheme as
follows: first, one sets the value of $\beta$ and calculates the
expectation value (or other moments) of the entropy at this $\beta$.
For small $N$, the expectations will be very close to their a priori
values due to the peakedness of ${\mathcal P}_{\beta}(S)$.
Afterwards, one integrates over $\beta(\xi)$ with the density
$\rho(\xi)$, which includes our a priori expectations about the
entropy of the distribution we are studying [${\mathcal
  P}\left(\beta\left(\xi\right)\right)$], as well as the evidence for
a particular value of $\beta$ [$\Gamma$-terms in Eq.~(\ref{rho})].

The crucial point is the behavior of the evidence. If it has a
pronounced peak at some $\beta_{\rm cl}$, then the integrals over
$\beta$ are dominated by the vicinity of the peak, $\widehat{S}$ is
close to $\xi(\beta_{\rm cl})$, and the variance of the estimator is
small. In other words, data ``selects'' some value of $\beta$, much in
the spirit of Refs.~\cite{mackay} -- \cite{nb}.  However, this
scenario may fail in two ways.  First, there may be no peak in the
evidence; this will result in a very wide posterior and poor
inference. Second, the posterior density may be dominated by $\beta
\to 0$, which corresponds to MLE, the best possible fit to the data,
and is a discrete analog of overfitting.  While all these situations
are possible, we claim that generically the evidence is well--behaved.
Indeed, while small $\beta$ increases the fit to the data, it also
increases the phase space volume of all allowed distributions and thus
decreases probability of each particular one [remember that $\langle
q_i \rangle_{\beta}$ has an extra $\beta$ counts in each bin, thus
distributions with $q_i < \beta/(N+\kappa)$ are strongly suppressed].
The fight between the ``goodness of fit'' and the phase space volume
should then result in some non--trivial $\beta_{cl}$, set by factors
$\propto N$ in the exponent of the integrand.

\begin{figure}[t]
  \begin{center}
    \begin{picture}(60,5)(0,0)
      \put(-60,0){(a)}
      \put(120,0){(b)}
    \end{picture}
  \end{center}
  \vspace{-1mm}
  \centerline{\epsfxsize=.49\hsize\epsffile{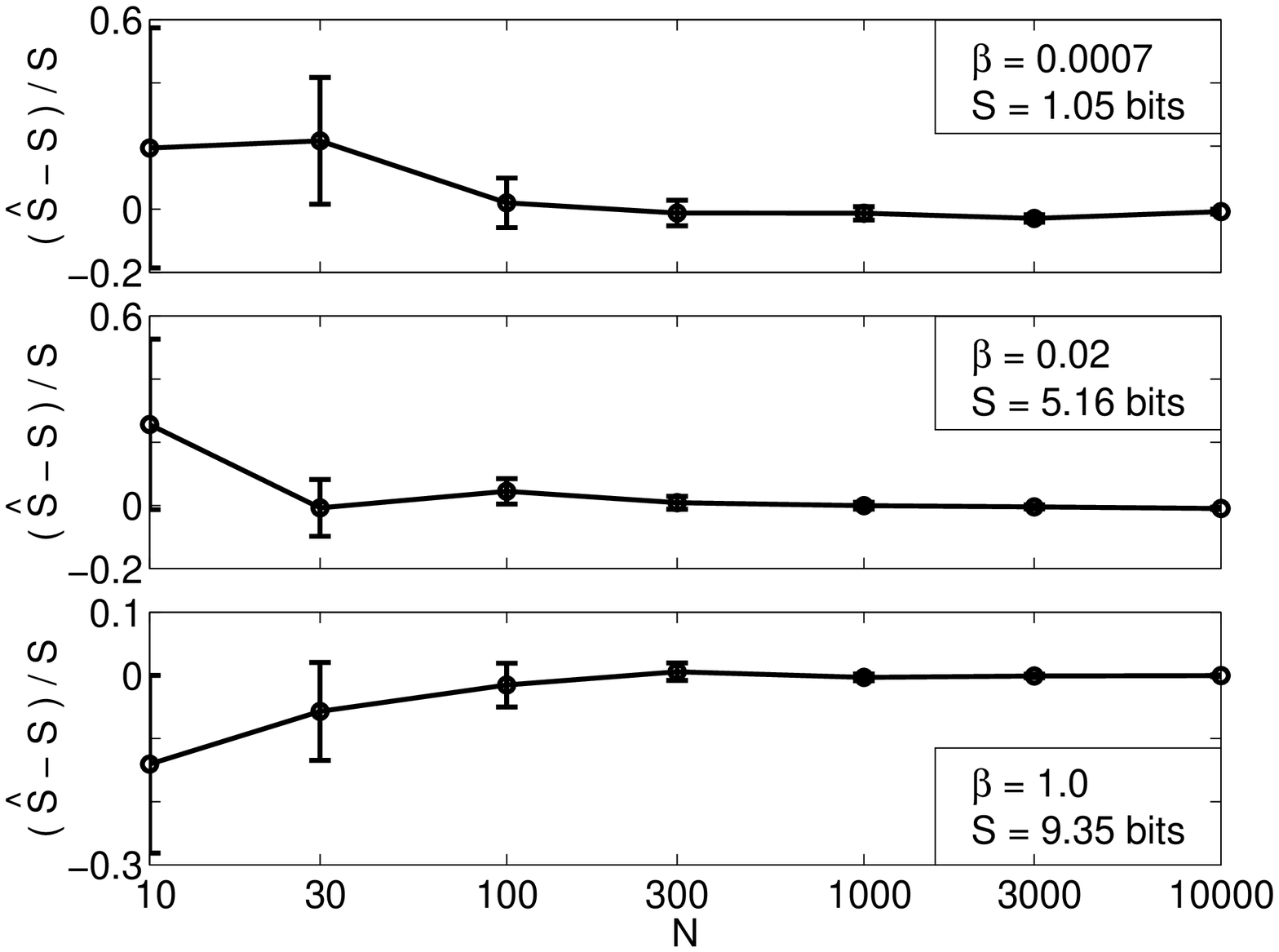}
    \epsfxsize=.49\hsize\epsffile{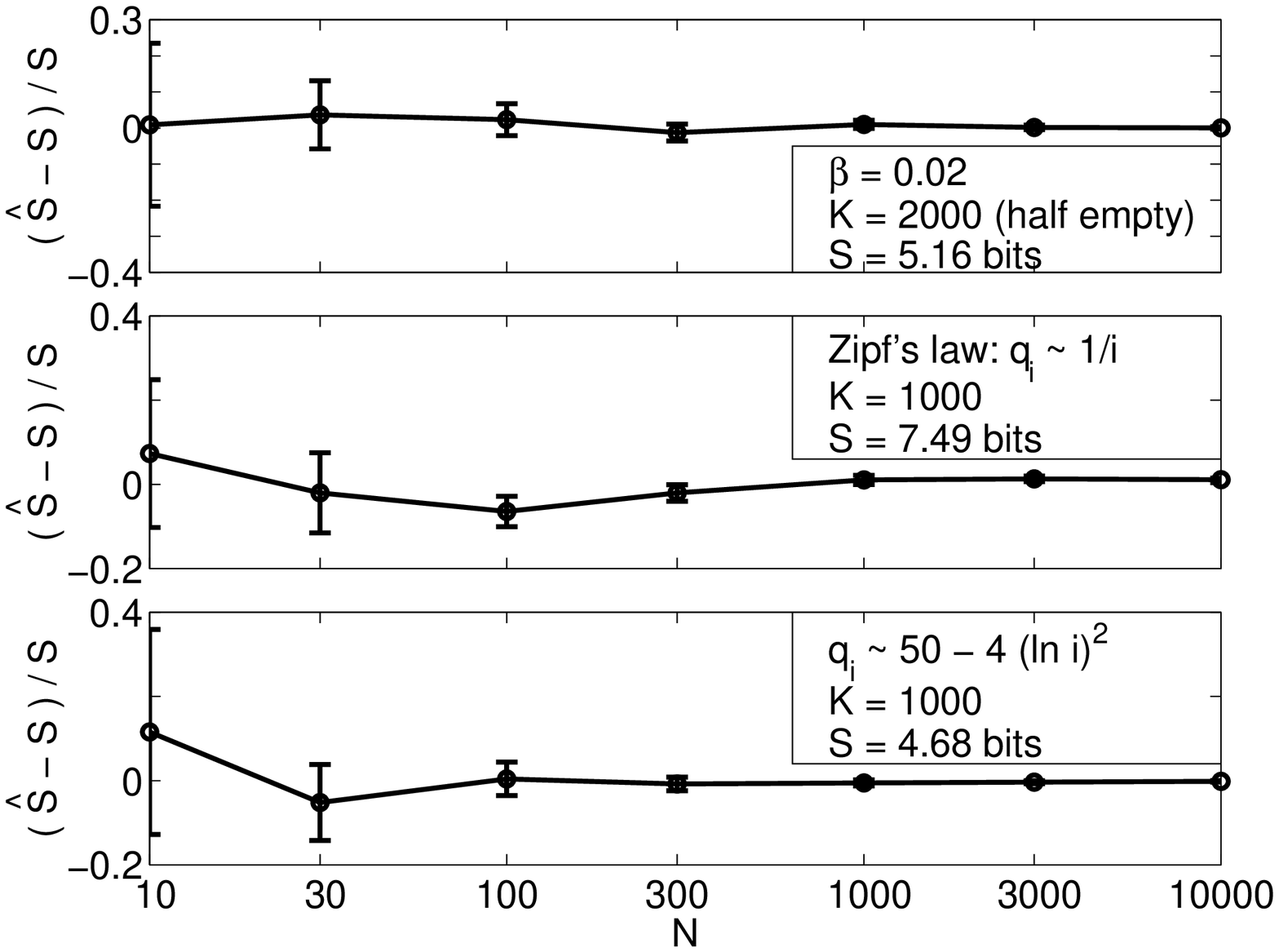}}
  \vspace{-4mm}
  \caption{Learning entropies with the prior Eq.~(\ref{Pflat}) and
    ${\mathcal P}(\beta)=1$. The actual relative errors of the
    estimator are plotted; the error bars are the relative widths of
    the posteriors. (a) Distributions from Fig.~\ref{example}. (b)
    Distributions atypical in the prior.  Note that while
    $\widehat{S}$ may be safely calculated as just $\langle S
    \rangle_{\beta_{\rm cl}}$, one has to do an honest integration
    over $\beta$ to get $\widehat{S^2}$ and the error bars.  Indeed,
    since ${\mathcal P}_{\beta} (S)$ is almost a $\delta$-function,
    the uncertainty at any fixed $\beta$ is very small (see
    Fig.~\ref{fixedbeta}).}
  \label{learning}
  \vspace{-4mm}
\end{figure}

Figure~\ref{learning} shows how the prior, Eq.~(\ref{Pflat}), performs
on some of the many distributions we tested. The left panel describes
learning of distributions that are typical in the prior ${\mathcal
  P}_{\beta}(\{q_i\})$ and, therefore, are also likely in ${\mathcal
  P}(\{q_i\};\beta)$. Thus we may expect a reasonable performance, but
the real results exceed all expectations: for all three cases, the
actual relative error drops to the $10\%$ level at $N$ as low as 30
(recall that $K=1000$, so we only have $\sim 0.03$ data points per bin
on average)! To put this in perspective, simple estimates like fixed
$\beta$ ones, MLE, and MLE corrected as in Eq.~(\ref{corr}) with $K^*$
equal to the number of nonzero $n_i$'s produce an error so big that it
puts them off the axes until $N >100$. \footnote{More work is needed to
  compare our estimator to more complex techniques, like in
  Ref.~\cite{srrb,pt}.}  Our results have two more nice features: the
estimator seems to know its error pretty well, and it is almost
completely unbiased.

One might be puzzled at how it is possible to estimate anything in a
1000--bin distribution with just a few samples: the distribution is
completely unspecified for low $N$! The point is that we are not
trying to learn the distribution --- in the absence of additional prior
information this would, indeed, take $N\gg K$ --- but to estimate
just one of its characteristics. It is less surprising that one number
can be learned well with only a handful of measurements. In practice
the algorithm builds its estimate based on the number of coinciding
samples (multiple coincidences are likely only for small $\beta$), as
in the  Ma's approach to entropy estimation from simulations of physical
systems
\cite{ma}.

What will happen if the algorithm is fed with data from a distribution
$\{\tilde{q}_i\}$ that is strongly atypical in ${\mathcal
  P}(\{q_i\};\beta)$? Since there is no $\{\tilde{q}_i\}$ in our
prior, its estimate may suffer.  Nonetheless, for any
$\{\tilde{q}_i\}$, there is some $\beta$ which produces distributions
with the same mean entropy as $S[\tilde{q}_i]$.  Such $\beta$ should
be determined in the usual fight between the ``goodness of fit'' and
the Occam factors, and the correct value of entropy will follow.
However, there will be an important distinction from the ``correct
prior'' cases. The value of $\beta$ indexes available phase space
volumes, and thus the smoothness (complexity) of the model class
\cite{bnt}. In the case of discrete distributions, smoothness is the
absence of high peaks. Thus data with faster decaying Zipf plots
(plots of bins' occupancy vs.\ occupancy rank $i$) are rougher. The priors ${\mathcal P}_{\beta}(\{q_i\})$ cannot account for all possible roughnesses. Indeed, they only generate distributions for which the expected number of bins $\nu$ with the probability mass less than some $q$ is given by $\nu(q) = K B(q, \beta, \kappa -\beta)$, where $B$ is the familiar incomplete Beta function, as in Eq.~(\ref{betadistr}). This means that the expected rank ordering for small and large ranks is
\begin{eqnarray}
q_i &\approx& 1 - \left[\frac{ \beta B(\beta, \kappa - \beta )  (K-1) \,i}
{K} \right] ^{1/(\kappa-\beta)}, \,\,\,\, i\ll K\,,
\label{left}\\
q_i &\approx& \left[ \frac{ \beta B(\beta, \kappa - \beta )  (K-i+1)}
{K}\right]^{1/\beta},\,\,\,\, K-i+1 \ll K\,.
\end{eqnarray}
In an undersampled regime we can observe only the first of the behaviors. Therefore, any
distribution with $q_i$ decaying
faster (rougher) or slower (smoother) than Eq.~(\ref{left}) for some $\beta$ cannot be explained
well with fixed $\beta_{\rm cl}$ for different $N$.  So, unlike in the cases of learning  data that are typical in ${\mathcal P}_{\beta}(\{q_i\})$, we should
expect to see $\beta_{\rm cl}$ growing (falling) for qualitatively
smoother (rougher) cases as $N$ grows.

\FORMAT{
\tabcolsep 0.5mm
\begin{wraptable}{r}{40.5mm}{
\begin{tabular}{cccc}
$N$  & 1/2 full & Zipf & rough\\ \hline
{\small units} & $\cdot 10^{-2}$ & $\cdot 10^{-1}$ & $\cdot 10^{-3}$ \\ \hline
10   & 1.7      & 1907 & 16.8\\
30   & 2.2      & 0.99 & 11.5\\
100  & 2.4      & 0.86 & 12.9\\
300  & 2.2      & 1.36 & 8.3 \\
1000 & 2.1      & 2.24 & 6.4 \\
3000 & 1.9      & 3.36 & 5.4 \\
10000& 2.0      & 4.89 & 4.5 \\
\end{tabular}}
\vspace{-3mm}
\caption{$\beta_{\rm cl}$ for solutions shown on Fig.~\ref{learning}(b).}
\label{betacl}
\end{wraptable}}{}{}

Figure~\ref{learning}(b) and Tbl.~\ref{betacl} illustrate these
points. First, we study the $\beta=0.02$ distribution from
Fig.~\ref{example}. However, we added a 1000 extra bins, each with
$q_i=0$.  Our estimator performs remarkably well, and $\beta_{\rm cl}$
does not drift because the ranking law remains the same. Then we turn
to the famous Zipf's distribution, so common in Nature. It has $n_i
\propto 1/i$, which is qualitatively smoother than our prior allows.
Correspondingly, we get an upwards drift in $\beta_{\rm cl}$. Finally,
we analyze a ``rough'' distribution, which has $q_i \propto 50 - 4(\ln
i)^2$, and $\beta_{\rm cl}$ drifts downwards. Clearly, one would want
to predict the dependence $\beta_{\rm cl}(N)$ analytically, but this
requires calculation of the predictive information (complexity) for the
involved distributions \cite{bnt} and is a work for the future. Notice that, the entropy estimator for atypical
\FORMAT{}{}{
\tabcolsep 0.5mm
\begin{wraptable}{r}{40.5mm}{
\begin{tabular}{cccc}
$N$  & 1/2 full & Zipf & rough\\ \hline
{\small units} & $\cdot 10^{-2}$ & $\cdot 10^{-1}$ & $\cdot 10^{-3}$ \\ \hline
10   & 1.7      & 1907 & 16.8\\
30   & 2.2      & 0.99 & 11.5\\
100  & 2.4      & 0.86 & 12.9\\
300  & 2.2      & 1.36 & 8.3 \\
1000 & 2.1      & 2.24 & 6.4 \\
3000 & 1.9      & 3.36 & 5.4 \\
10000& 2.0      & 4.89 & 4.5 \\
\end{tabular}}
\vspace{-3mm}
\caption{$\beta_{\rm cl}$ for solutions shown on Fig.~\ref{learning}(b).}
\label{betacl}
\FORMAT{}{}{\vspace{-3mm}}
\end{wraptable}}
 cases is almost as
good as for typical ones.  A possible exception is the 100--1000
points for the Zipf distribution---they are about two standard
deviations off. We saw similar effects in some other ``smooth'' cases
also.  This may be another manifestation of an observation made in
Ref.~\cite{nb}: smooth priors can easily adapt to rough distribution,
but there is a limit to the smoothness beyond which rough priors
become inaccurate.

To summarize, an analysis of a priori entropy statistics in common
power--law Bayesian estimators revealed some very undesirable features. We are fortunate, however, that these minuses can be easily
turned into pluses, and the resulting estimator of entropy is precise,
knows its own error, and gives amazing results for a very large class of
distributions.

\section*{Acknowledgements}
We thank Vijay Balasubramanian, Curtis Callan, Adrienne Fairhall, Tim
Holy, Jonathan Miller, Vipul Periwal, Steve Strong, and Naftali Tishby for useful
discussions. I.\ N.\ was supported in part by NSF Grant No.\ PHY99-07949 to the Institute for Theoretical Physics.


\begin{thebibliography}{99}
\itemsep 0mm
{\small
    \bibitem{mackay}\newblock{D.~MacKay, {\it Neural Comp.} {\bf 4},
    415--448 (1992).}

    \bibitem{vijay}\newblock{V.~Balasubramanian, {\em Neural Comp.}
    {\bf 9}, 349--368 (1997)\FORMAT{.}{, {\tt \small
        adap-org/9601001}.}{, {\tt \small adap-org/9601001}.}}

    \bibitem{bcs}\newblock{W.~Bialek, C.~Callan, and S.~Strong, {\it
      Phys.~Rev.~Lett.}  {\bf 77}, 4693--4697 (1996)\FORMAT{.}{, {\tt
        \small cond-mat/9607180}.}{, {\tt \small cond-mat/9607180}.}}

    \bibitem{nb}\newblock{I.~Nemenman and W.~Bialek, {\it Advances in
      Neural Inf.\ Processing Systems} {\bf 13}, 287--293 (2001)\FORMAT{.}{,
      {\tt \small cond-mat/0009165}.}{, {\tt \small
        cond-mat/0009165}.}}

    \bibitem{maxent}\newblock{J.~Skilling, in {\it Maximum entropy and
      Bayesian methods,} J.~Skilling ed. (Kluwer Academic Publ.,
    Amsterdam, 1989), pp.~45--52.}

    \bibitem{ww}\newblock{D.~Wolpert and D.~Wolf, {\it Phys.~Rev.~E}
    {\bf 52}, 6841--6854 (1995)\FORMAT{.}{, {\tt \small
        comp-gas/9403001}.}{, {\tt \small comp-gas/9403001}.}}

    \bibitem{thesis}\newblock{I.~Nemenman, Ph.D. Thesis, Princeton,
    (2000), ch.~3, \FORMAT{\small
      http://arXiv.org/abs/physics/0009032} {\tt \small
      physics/0009032} {\tt \small physics/0009032}.}

\bibitem{laplace}\newblock{P.~de Laplace, marquis de, {\em Essai philosophique sur les probabilit\'es} (Courcier, Paris, 1814), trans.\ by F.~Truscott and F.~Emory, {\em A philosophical essay on probabilities}  (Dover, New York, 1951).}

\bibitem{hardy}\newblock{G.~Hardy, {\em Insurance Record} (1889), reprinted in {\em Trans.~Fac.~Actuaries} {\bf 8} (1920).}

\bibitem{lidstone}\newblock{G.~Lidstone, {\em Trans.~Fac.~Actuaries} {\bf 8}, 182--192 (1920).}

\bibitem{jeffreys}\newblock{H.~Jeffreys, {\em Proc.~Roy.~Soc.~(London) A} {\bf 186}, 453--461 (1946).} 

\bibitem{kt}\newblock{R.~Krichevskii and V.~Trofimov, {\em IEEE Trans.\ Inf.\ Thy.} {\bf  27}, 199--207 (1981).}

    \bibitem{wst}\newblock{F.~Willems, Y.~Shtarkov, and T.~Tjalkens,
    {\it IEEE Trans.\ Inf.\ Thy.} {\bf 41}, 653--664 (1995).}

    \bibitem{sg}\newblock{T.~Schurmann and P.~Grassberger, {\it Chaos}
    {\bf 6}, 414--427 (1996).}

    \bibitem{srrb}\newblock{S.~Strong, R.\ Koberle, R.\ de Ruyter van Steveninck, and W.\ Bialek, {\em Phys.\ Rev.\ Lett.}
    {\bf 80}, 197--200 (1998)\FORMAT{.}{, {\tt \small
        cond-mat/9603127}.}{, {\tt \small cond-mat/9603127}.}}

    \bibitem{pt}\newblock{S.~Panzeri and A.~Treves, {\em Network:
      Comput. in Neural Syst.} {\bf 7}, 87--107 (1996).}

\bibitem{mixt}\newblock{K.\ Sjšlander, K.\ Karplus, M.\ Brown, R.\ Hughey, A.\ Krogh, I. S.\ Mian, and D.\ Haussler, 
{\em Computer Applications in the Biosciences (CABIOS)} {\bf 12}, 327--345 (1996).}

    \bibitem{ma}\newblock{S.~Ma, {\em J.\ Stat.\ Phys.} {\bf 26}, 221
    (1981).}

    \bibitem{bnt}\newblock{W.~Bialek, I.~Nemenman, N.~Tishby, {\em Neural Comp.} {\bf 13}, 2409-2463 (2001)\FORMAT{.}{, {\tt
        \small physics/0007070}.}{, {\tt \small physics/0007070}.}}  }

\end{thebibliography}
\end{document}